\documentstyle[preprint,aps,epsf]{revtex}
\newcommand{\be}{\begin{equation}}
\newcommand{\ee}{\end{equation}}
\newcommand{\bea}{\begin{eqnarray}}
\newcommand{\eea}{\end{eqnarray}}

\begin{document}
\preprint{SU-ITP-97-22}
\title{5d Field Theories and M Theory}
\author{\bf Barak Kol}
\address{ Department of Physics,\\ Stanford University,\\ Stanford, CA
94305, USA}
\address{barak@leland.stanford.edu}
\maketitle
\begin{abstract}
5-brane configurations describing 5d field theories are promoted to an M theory description \`{a} la Witten in terms of polynomials in two complex variables. The coefficients of the polynomials are the Coulomb branch. This picture resolves apparent singularities  at vertices and reveals exponentially small corrections. These corrections ask to be compared to world line instanton corrections. From a different perspective this procedure may be used to define a diagrammatic representation of polynomials in two variables.
\end{abstract}

\newpage
\bigskip

\section {The IIB construction}
5d N=1 field theories were constructed from configurations of (p,q) 5-branes by Amihay Hanany and Ofer Aharony \cite{AH}, generalizing the configurations of Hanany-Witten \cite{HW}. The essential geometry takes place in a plane which they denote $(x_3,x_7)$, and will be denoted here by $(x,y)$. Fig 2 is a typical example. Each line represents a (p,q) 5-brane (5B) of IIB string theory with tension 

\be
T^2 \sim \left({p \over \lambda}\right)^2 + \left( {q \over \lambda ^2}\right) ^2
\label{tension}
\ee
where $\lambda$ is the IIB string coupling. Not shown are 4+1 additional dimensions shared by all 5Bs. Charge conservation at a vertex requires :

\be
{\sum_{i}^{}{p_i}} = {\sum_{i}^{}{q_i}} = 0.
\label{vcharge}
\ee

We also have to satisfy the no-force condition at the vertex. By constraining the slope of any line representing a (p,q) 5B through

\be 
\Delta x : \Delta y = p : q/\lambda
\label{v1}
\ee
the no-force condition is a consequence of eq's (\ref{tension},\ref{vcharge}). Actually it was shown in \cite{AH} that this condition preserves one quarter of the supersymmetries. Thus we are discussing theories with 8 susy's (corresponding to minimal N=1 in 6d and 5d and to N=2 in 4d). With these constraints, the 5DBs, (p,q)=(1,0), are horizontal, and the NS 5Bs, (p,q)= (0,1),  are vertical.

The transition to a 5d field theory from this configuration proceeds as follows : Assuming small $\lambda$ , the low energy theory is a 6d field theory describing the light 5DB  supported by heavy NS 5B. Furthermore, due to the finite extent of the 5DBs in the $x$ direction, at low enough energy the field theory loses a dimension and becomes 5d. 
%%%%%%%%%%%%%%%%%%%%%%%%%%%%%%%%%%%%%%%%%%%%

\section {M theory Origin}
Following ``Solutions of Four Dimensional Field Theories Via M Theory'' /Witten \cite{W}, an M theory description will be sought.
Type IIB string theory can be reached by compactifying M theory on a small torus \cite{Sch} of size $L_1$ times $L_2$. The relations between the parameters of the two theories are 

\bea
\lambda = {L_1 \over L_2}&&  \qquad
{l_s}^2 = {  {l_{11}}^3 \over  L_1} \nonumber \\
L_B &&= { {l_{11}}^3 \over {L_1 L_2} }
\label{2BM}
\eea
where $l_s$ is the length scale of the fundamental string, and $L_B$ is the size of the ``new'' IIB dimension. (Warning : the preceding equations (\ref{2BM}) hold only up to constants).

A 5B of IIB has two possible M theory origins : a 5MB or  a 
wrapped KK monopole. Following \cite{W} a 5MB configuration is chosen. A (p,q) 5B is a 5MB wrapping a (p,q) cycle on the torus and regenerating an extra world  volume dimension by wrapping $L_B$. Thus in M theory the configuration is described by a single 5MB. The vertex singularities at the intersection of two (p,q) 5Bs are expected to be smoothed out.
%%%%%%%%%%%%%%%%%%%%%%%%%%%%%%%%%%%%%%%%%%%%

\section{Complex Variables}
Denote the coordinates on the torus by $(x_t,y_t)$. Now the slope condition is that the slope in the $(x,y)$ plane equals the one in $(x_t,y_t)$
\be
\Delta x : \Delta y = \Delta x_t : \Delta y_t = p : q/\lambda.
\ee
This condition can be written more compactly by introducing complex coordinates 

\be
w = x + i x_t  \qquad z = y + i y_t 
\label{wz}
\ee
and dimensionless coordinates

\be
\tilde{w} = w/ {L_1} = \tilde{x} + i \tilde{x_t}  \qquad \tilde{z} = z/ {L_2} = \tilde{y} + i \tilde{y_t}  
\label{tilde}
\ee
Now the no-force condition (\ref{v1}) simplifies to

\be
\Delta \tilde{w} : \Delta \tilde{z} = p : q
\label{v2}
\ee

In M theory the no-force BPS condition translates \cite{W,BckS} to the "supersymmetric cycle" condition, requiring that the 5B be described by a complex curve. Since $\tilde{w}, \tilde{z}$ live on a cylinder, define single valued complex coordinates

\be
s = exp (\tilde{w})  \qquad t = exp (\tilde{z})
\label{st}
\ee

Finally, the 5MB configuration is defined by a curve

\be
F (s,t) = 0.
\ee
%%%%%%%%%%%%%%%%%%%%%%%%%%%%%%%%%%%%%%%%%%

\section{Example 1}
Consider the IIB configuration in figure 1a. 

This is the simplest vertex. I will construct the curve F by analyzing the limits :
\begin{enumerate}
\item  As $s\rightarrow 0$  $\left | t \right | =1 \Rightarrow F = t - 1 + O(s).$
\item As $t\rightarrow 0$  $\left | s \right | =1 \Rightarrow F = s - 1 + O(t).$
\item As  $s,t \rightarrow \infty \ s \sim t.$
\end{enumerate}

All these conditions are met by

\be
F (s,t) = s + t -1
\label{F1}
\ee

The smoothed out configuration, projected on $(\tilde{x},\tilde{y})$ is given in figure 1b.

Analyze the corrections. Consider the envelope of the smooth configuration (1b). For instance, the lower left curve is given by

\be
exp (\tilde{x}) + exp (\tilde{y}) - 1 =0.
\label{envelope1}
\ee

We see that near the vertex there are order 1 corrections, and away from it, the envelope approaches the original lines (1a) up to exponentially small terms

\be
exp (- \left | \tilde{x} \right | ) \ \ {\rm and}  \qquad exp (- \left | \tilde{y} \right | )
\label{cor}
\ee

Since the correction is at most order 1 in the dimensionless variables, the dimensionful corrections vanish with the size of the torus.

Analyze the Coulomb branch. As in \cite{W,AH}, it is given by the parameters of the polynomial  that do not change the asymptotic configuration. An overall multiplicative parameter in F is irrelevant. Translating x,y corresponds to a (complex) rescaling of $s,t$, accounting for the two other parameters. Thus the Coulomb branch is null.
%%%%%%%%%%%%%%%%%%%%%%%%%%%%%%%%%%%%%%%%%%%

\section{Example 2}

Consider the brane configuration in figure 2. It is the simplest field theory configuration resulting in an SU(2) gauge theory in 5d N=1. The parameters of the gauge theory can be read off from the geometry of the configuration. The vertical separation between the 5DBs , $\Delta y$, is a flat direction proportional to $m_W$, the mass of the W boson
\bea
m_W =&& {\Delta y \over {l_s}^2} \nonumber \\
\Delta \tilde{y} =&& {\Delta y \over L_2} = {{m_W {l_{11}}^3} \over {L_1 L_2}} = m_W L_4
\label{mw}
\eea
When this distance is taken to vanish, the horizontal distance between the NS 5B reduces accordingly and becomes proportional to $1/g^2$, where $g$ is the gauge coupling 
\bea
{1 \over g^2} =&& {\Delta x \over {\lambda {l_s}^2} } = { \Delta x L_2 \over {l_{11}}^3} \nonumber \\
\Delta \tilde{x} =&& {\Delta x \over L_1} =  {L_4 \over g^2}
\label{g}
\eea
The area enclosed by the central rectangle is the tension of the monopole (a string in 5d). As in \cite{W}, the shape of the NS 5B can be identified with the running coupling constant.

To determine the polynomial, I use the same method as in the previous section and find

\be
F (s,t) = t (s^2 +1) + A s ( t^2 + 1) + A B s t
\ee
The picture after resolving the singularities looks like the original picture, with each vertex replaced by the resolved vertex (figure 1b).

{\bf The corrections.} Using dimensionless variables the corrections decrease exponentially  with the distance from the vertices. From eq's (\ref{mw},\ref{g}) we see that for fixed parameters of the field theory, these distances are proportional to $L_4$. Thus the corrections are of the form

\be 
exp (L_4 p)
\label{cor2}
\ee
where p is some momentum scale such as $m_W$ or ${1 \over g^2}$. This expression resembles expected field theoretic corrections for a theory on a finite radius $L_4$ attributed to world line instantons. The field theory correction to $m_W$, to the mass of the monopole, or to the runninng coupling constant could be calculated and compared with the geometrical correction.
 
The Coulomb branch. The polynomial has 5 complex coefficients. As before three get fixed. The parameter A changes the asymptotic configuration and is related to the coupling constant. The parameter B is the complex modulus of the theory. 
%%%%%%%%%%%%%%%%%%%%%%%%%%%%%%%%%%%%%%%%%%

\section{Diagrams for Polynomials}

The preceding relation of polynomials to diagrams might be studied for its own sake. It seems that for a given diagram  we can associate a polynomial $F (s,t)$ in 2 complex variables, or  to be more precise, a limit of polymials as $L_4 \rightarrow \infty$. Reversing the process, given a polynomial we can drive it to its limit, and associate a diagram to it.  The diagram consists of the projection of its zero surface $F=0$ on the $(\hat{x},\hat{y})$ plane, where
\be
\hat{x}=Re (log (s) )/ L_4 \qquad \hat{y} = Re( log(t))/ L_4
\ee
The diagram is composed of lines, each carrying a (p,q) index, and vertices which conserve the (p,q) charge. It remains to be proved that the diagrams always satisfy these rules, and to check when two polynomials give the same diagram.

\begin{center}
{\bf ACKNOWLEDGEMENTS}
\end{center}
I thank A. Hanany for introducing me to the subject and I thank L. Susskind,  J. Rahmfeld, A. Rajaraman and W. K. Wong. This work is supported by NSF grant PHY-9219345. 

\begin{figure}
\centerline{\epsfxsize=70mm\epsfbox{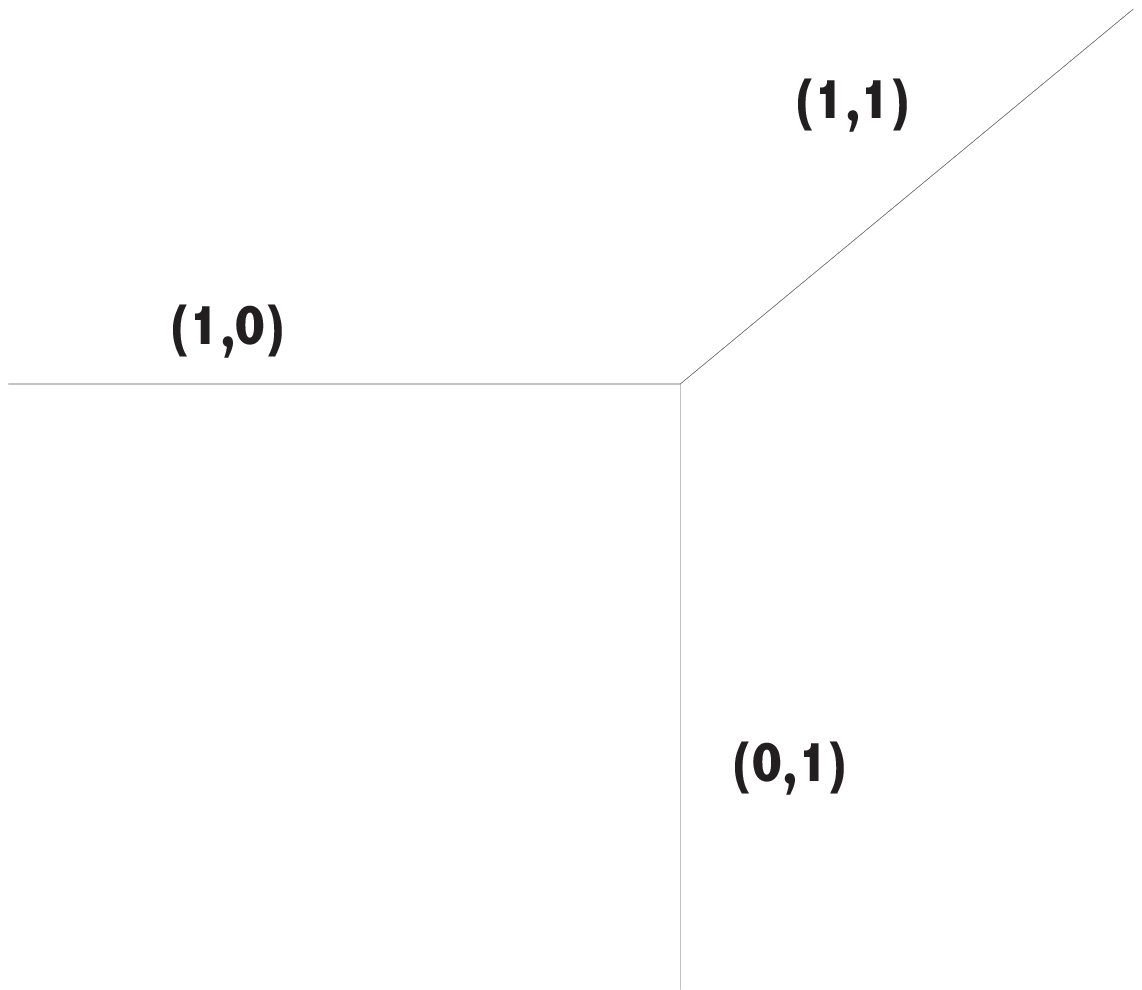}\epsfxsize=70mm
\epsfbox{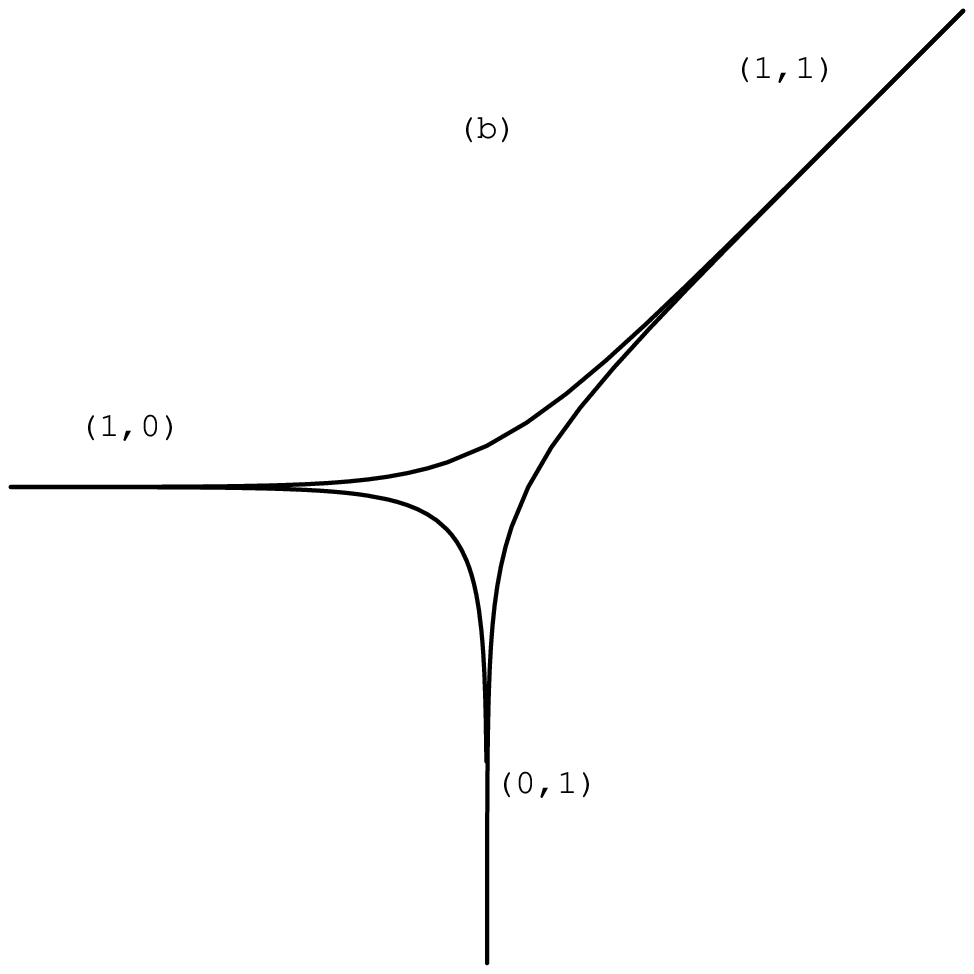}}
\medskip
\caption{A vertex configuration (a - left) is smoothed out in (b -right)}
\end{figure}
\begin{figure}
\centerline{\epsfxsize=80mm\epsfbox{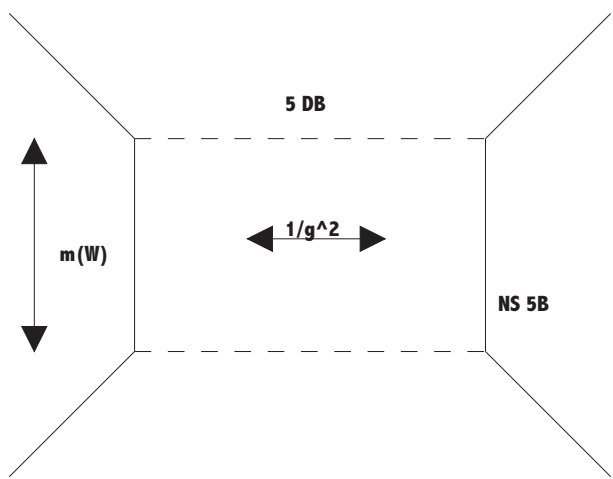}}
\medskip
\caption{A 5brane configuration resulting in an SU(2) 5d theory}
\end{figure}

\end{document}